\documentclass[12pt]{amsart}

\newtheorem{thm}{Theorem}
\newtheorem{prop}{Proposition}
\newtheorem{cor}{Corollary}

\newcommand{\cset}{\mathbf{C}} 
\newcommand{\rset}{\mathbf{R}} 
\newcommand{\maF}{\mathcal{F}}
\newcommand{\maD}{\mathcal{D}} 
\newcommand{\tix}{{\tilde x_h}}

\begin{document}

\title[The dequantization transform]%
{The dequantization transform and generalized Newton polytopes}

\author{G.~L.~Litvinov}
\address{Independent University of Moscow\\ 
11 Bol'shoi Vlasievskii per.\\
Moscow 121002, Russia }
\email{islc@dol.ru} 

\author{G.~B.~Shpiz}
\address{Moscow Center for Continuous Mathematical Education\\ 
11 Bol'shoi Vlasievskii per.\\
Moscow 121002, Russia } 
 
\thanks{This work has been supported by the RFBR grant 02--01--01062
and the Erwin Schr\"odinger International Institute for
Mathematical Physics (ESI)}

\begin{abstract}
For functions defined on $\cset^n$ or $\rset^n_+$ we construct a dequantization 
transform $f \mapsto\hat f$; this transform is closely related to the Maslov 
dequantization. 
If $f$ is a polynomial, then the sub\-differential $\partial\hat f$ of $\hat f$ 
at the origin
coincides 
with the Newton polytope of $f$. For the semiring of polynomials with 
nonnegative coefficients, the dequantization transform is a homomorphism of this 
semiring to the idempotent semiring of convex polytopes with respect to the 
well-known Minkowski operations. Using the dequantization transform we generalize 
these results to a wide class of functions and convex sets.
\end{abstract}

\maketitle

\section{Introduction. The Maslov dequantization and the dequantization transform}

Let $\rset$ and $\cset$ be the fields of real and complex numbers. The 
well-known max-plus algebra $\rset_{\max}= \rset\cup\{-\infty\}$ is defined by
the operations 
$x\oplus y=\max\{x, y\}$ and $x\odot y= x+y$. 

The max-plus algebra can be 
treated as a result of the 
Maslov dequantization of the semifield $\rset_+$ of all nonnegative numbers, 
see, 
e.g., \cite{Mas87, LiMa95}. The change of variables
\begin{equation}
x\mapsto u=h\log x,
\end{equation}
where $h>0$, defines a map $\Phi_h\colon \rset_+\to \rset\cup\{-\infty\}$. Let the 
addition 
and multiplication operations be mapped from 
$\rset_+$ to $\rset\cup\{-\infty\}$ by 
$\Phi_h$, i.e.\ let 
\begin{gather*}
u\oplus_h v = h \log(\exp(u/h)+\exp(v/h)),\quad u\odot v= u+ v,\\ 
\mathbf{0}=-\infty = \Phi_h(0),\quad \mathbf{1}= 0 = \Phi_h(1). 
\end{gather*}
It can easily be checked that 
$u\oplus_h v\to \max\{u, v\}$ as $h\to 0$. Thus we get the semifield 
$\rset_{\max}$ (i.e.\ the max-plus algebra) with zero $\mathbf{0}= -\infty$ and unit 
$\mathbf{1}=0$ as a result of this deformation of the algebraic structure in 
$\rset_+$. 

The semifield $\rset_{\max}$ is a typical example of an {\it
 idempotent semiring}; this is a semiring with idempotent addition, i.e.,
 $x\oplus x = x$ for arbitrary element
 $x$ of this semiring, see, e.g., \cite{Gun98}.

The analogy with quantization is obvious; the parameter $h$ plays the role of 
the Planck constant \cite{LiMa95}. In fact the Maslov dequantization is the usual Schr\"odinger dequantization but for pure imaginary values of the Planck 
constant (see, e.g., \cite{LiMaSh2001}). The map $x\mapsto|x|$ and the 
Maslov dequantization for $\rset_+$ give us a natural passage from the field 
$\cset$ (or $\rset$) to the max-plus algebra $\rset_{\max}$.

Let $X$ be a topological space. For functions $f(x)$ defined on $X$
we shall say that a certain property is valid {\it almost everywhere} (a.e.) if 
it is valid for all elements $x$ of an open dense subset of $X$. 

Suppose $X$ is $\cset^n$ or $\rset^n$; denote by $\rset^n_+$ the set
$x=\{\,(x_1, \dots, x_n)\in X \mid x_i\geq 0 \text{ for $i = 1, 2, \dots, n$}\,\}$.
 For $x= (x_1, \dots, x_n) \in X$ we set 
$\exp(x) = (\exp(x_1), \dots,\linebreak \exp(x_n))$;
so if $x\in\rset^n$, then $\exp(x)\in \rset^n_+$. 

Denote by $\maF(\cset^n)$ the set of all functions defined
and continuous on an open dense subset $U\subset \cset^n$
such that $U\supset \rset^n_+$. In all the examples below we consider
even more regular functions, which are holomorphic in~$U$.
 It is clear that $\maF(\cset^n)$ is a ring 
(and an algebra over $\cset$) with respect to the usual addition
and multiplications of functions.

For $f\in \maF(\cset^n)$ let us define the function $\hat f_h$
by the following formula:
\begin{equation}
\hat f_h(x) = h \log|f(\exp(x/h))|,
\end{equation}
where $h$ is a (small) real parameter and $x\in\rset^n$. Set
\begin{equation}
\hat f(x) = \lim_{h\to 0} \hat f_h (x),
\end{equation}
if the right-hand part of (3) exists almost everywhere. We shall say that the 
function $\hat f(x)$ is a {\it dequantization} of the function $f(x)$ and the 
map $f(x)\mapsto \hat f(x)$ is a {\it dequantization transform}. By 
construction, $\hat f_h(x)$ and $\hat f(x)$ can be treated as functions taking their 
values in $\rset_{\max}$. Note that in fact $\hat f_h(x)$ and $\hat f(x)$
 depend on the restriction of $f$ to $\rset_+^n $
only; so in fact the dequantization transform is constructed
for functions defined on $\rset^n_+$ only. 
It is clear that the dequantization transform is 
generated by the Maslov dequantization and the map $x\mapsto |x|$. Of 
course, similar definitions can be given for functions defined on $\rset^n$
and $\rset_+^n$.

We shall see that if $f(x)$ is a polynomial, then there exists the 
dequantization $\hat f$ of this polynomial and the subdifferential 
$\partial \hat f$ of the function $\hat f$ coincides with the Newton polytope 
of the polynomial $f$.

It is well known that all the convex compact subsets in $\rset^n$ form an 
idempotent 
semiring $\mathcal{S}$ with respect to the Minkowski operations: for $A, B \in \mathcal{S}$ 
the sum $A\oplus B$ is the convex hull of the union $A\cup B$; the product 
$A\odot B$ is defined in the following way: $A\odot B = \{\, x\mid x = a+b, 
\text{ where $a\in A, b\in B$}\,\}$. In fact $\mathcal{S}$ is an idempotent linear space over
 $\rset_{\max}$ (see, e.g., \cite{LiMaSh2001}). Of course, the Newton polytopes 
in $V$ form a 
subsemiring $\mathcal{N}$ in $\mathcal{S}$. If $f$, $g$ are polynomials, then 
$\partial(\widehat{fg}) = \partial\hat f\odot\partial\widehat g$; moreover, if $f$ and $g$ are 
``in general position'', then $\partial(\widehat{f+g}) = \partial\hat 
f\oplus\partial\widehat g$. For the semiring of all polynomials with 
nonnegative coefficients the dequantization transform is a homomorphism of this 
``traditional'' semiring to the idempotent semiring $\mathcal{N}$.

Our aim is to prove these results and to generalize them to a more extensive 
class of functions and convex sets.

The authors are grateful to A.~N.~Sobolevski{\u\i} for his help and useful discussions.

\section{The dequantization transform: algebraic properties}
 
Denote by $V$ the set $\rset^n$ treated as a linear Euclidean
space (with the scalar product $(x, y) = x_1y_1+ x_2y_2 +\dots + x_ny_n$)
and set $V_+ = \rset_+^n$.
We shall say that a function $f\in \maF(\cset^n)$ is {\it dequantizable} 
whenever its 
dequantization $\hat f(x)$ exists (and is defined on an open dense subset 
of $V$). By $\maD (\cset^n)$ denote the set of all dequantizable functions and by 
$\widehat{\maD}(V)$ denote the set $\{\,\hat f \mid f\in \maD(\cset^n)\,\}$. Recall that 
functions from $\maD(\cset^n)$ (and $\widehat{\maD}(V)$) are defined almost everywhere and 
$f=g$ means that $f(x) = g(x)$ a.e., i.e., for $x$ ranging over an open dense subset 
of $\cset^n$ (resp., of $V$). Denote by $\maD_+(\cset^n)$ the set of all 
functions $f\in \maD(\cset^n)$ 
such that $f(x_1, \dots, x_n)\geq 0$ if $x_i\geq 0$ for $i= 1,\dots, n$; so
 $f\in \maD_+(\cset^n)$ if the restriction of $f$ to $V_+ = \rset_+^n$ is a
 nonnegative function. By $\widehat{\maD}_+(V)$ denote the image of $\maD_+(\cset^n)$
 under the dequantization transform. We shall say that functions 
$f, g\in \maD(\cset^n)$ are in {\it general position} whenever
$\hat f (x) \neq \widehat g(x)$ for $x$ running an open dense
subset of $V$.

\begin{thm}
For functions $f, g \in \maD(\cset^n)$ and any nonzero constant $c$, the 
following equations are valid:

\begin{enumerate}
\item[1)] $\widehat{fg} = \hat f + \widehat g$;
\item[2)] $|\hat f| = \hat f$; $\widehat{cf} = f$; $\widehat c =0$;
\item[3)] $(\widehat{f+g})(x) = \max\{\hat f(x), \widehat g(x)\}$ a.e.\ if $f$ and $g$
are nonnegative on $V_+$ (i.e., $f, g \in \maD_+(\cset^n)$) or $f$ and $g$
 are in general position.
\end{enumerate}
Left-hand sides of these equations are well-defined automatically.
\end{thm}

\begin{proof} Statements 1) and 2) can be easily deduced from our basic
 definitions and formulas (2) and (3).

 Let us prove statement 3). Set $\tix = \exp(x/h) \in V_+$.
 Suppose $f, g \in \maD_+(\cset^n)$; then $|f| = f$, $|g| = g$,
 $|f+g| = f+ g$ on $V_+$ and we have the following inequalities:
\[\max \{ f (\tix), g (\tix)\} \leq (f+g) (\tix) \leq 2
 \max \{ f (\tix), g (\tix)\}.\] 
Hence 
\[h \log (\max \{ f (\tix), g (\tix)\}) 
\leq h\log((f+ g) ( \tix)) \leq h \log 2 +
 h \log (\max \{ f (\tix), g(\tix)\}).\]  
 But $h\log 2\to 0$ as $h\to 0$,
 and the logarithmic function is monotonic.    
Thus we have
\[\max\{ \hat f (x), \widehat g (x)\} \leq \widehat{(f+g)} (x) \leq \max \{\hat f (x),
 \widehat g (x)\},\] 
which completes this part of our proof. 

A similar idea can
 be used if functions $f$ and $g$ are in general position. Without loss of
 generality we can suppose that $\hat f (x) < \widehat g (x)$ almost
 everywhere in~$V$.  Take an $x\in V$ where this inequality holds.
In this case there exists a positive number $c$ such that 
$h\log|f(\tix)| + c < h \log|g(\tix)|$ if
 the parameter $h$ is small enough.         
 Hence $|f(\tix)|
 \exp (c/h) < |g(\tix)|$. Note that $\exp(c/h)\to \infty$ as
 $h\to \infty$; hence $\exp (c/h) > 2$ if $h$ is small enough.
 Therefore we have $|f(\tix))| < (1/2) |g(\tix)|$
 and $(1/2)|g(\tix)| < |f(\tix) + g(\tix)|$. 
On the other hand, we obviously have the inequality $|(f + g)(\tix)| <
 2 |g(\tix)|$. So we get 
\[\frac12|g(\tix)| < |(f + g)(\tix)| < 2
 |g(\tix)|;\] 
from this and from formulas (2) and (3) it follows that
$$
\widehat g(x) \leq \widehat{(f + g)} (x) \leq \widehat g (x) = \max \{ \hat f (x), \widehat g (x)\}.
$$
This concludes the proof.
\end{proof}

\begin{cor}
The set $\maD_+(\cset^n)$ has a natural structure of a semiring with respect to 
the
 usual addition and multiplication of functions taking their values in $\cset$.
 The set $\widehat{\maD}_+(V)$ has a natural
 structure of an idempotent semiring with respect to the operations 
$(f\oplus g)(x) = \max \{ f(x), g(x)\}$, $(f\odot g)(x) = f(x) + g(x)$; 
elements
 of $\widehat{\maD}_+(V)$ can be naturally treated as functions taking their values
 in $\rset_{\max}$. The dequantization transform generates a homomorphism from
 $\maD_+(\cset^n)$ to $\widehat{\maD}_+(V)$.
\end {cor}

\section{Generalized polynomials and simple functions}

For any nonzero number $a\in\cset$ and any vector
 $d = (d_1, \dots, d_n)\in V = \rset^n$ 
we set $m_{a,d}(x) = a \prod_{i=1}^n x_i^{d_i}$; functions of this kind we
 shall 
call {\it generalized monomials}. Generalized monomials are defined a.e.\ on
$\cset^n$ and on $V_+$, but not on $V$ unless the numbers $d_i$ take
integer or suitable rational values. We shall say that a function $f$ is a {\it 
generalized polynomial} whenever it is a finite sum of linearly
 independent generalized monomials. For instance, Laurent polynomials
are examples of generalized polynomials.

As usual, for $x, y\in V$ we set $(x,y) = x_1y_1 + \dots + x_ny_n$. The 
following proposition is a result of a trivial calculation.

\begin{prop}
For any nonzero number $a\in V = \cset$ and any vector $d\in V = \rset^n$ 
we have $(\widehat{m_{a,d}})_h(x) = (d, x) + h\log|a|$.
\end{prop}

\begin{cor}
If $f$ is a generalized monomial, then $\hat f$ is a linear function.
\end{cor}

Recall that a real function $p$ defined on $V = \rset^n$ is {\it sublinear} 
if $p = \sup_{\alpha} 
p_{\alpha}$, where $\{p_{\alpha}\}$ is a collection of linear functions. 
Sublinear functions defined everywhere on $V=\rset^n$ are convex; thus these
 functions are continuous, see \cite{Ro70}, Theorem~5.5 and Corollary~10.1.1.
 We discuss sublinear functions of this kind only. Suppose $p$ is a continuous
 function defined on $V$, then $p$ is sublinear whenever 

1) $p(x+ y) \leq p(x) + p(y)$ for all $x, y \in V$;

2) $p(cx) = cp(x)$ for all $x\in V$, $c\in \rset_+$.

So if $p_1$, $p_2$ are sublinear functions, then $p_1 +p_2$ is a sublinear
 function.

We shall say 
that a function $f \in \maF(\cset^n)$ is {\it simple}, if its dequantization $\hat f$ 
exists and a.e.\ coincides with a sublinear function; by misuse of language, we 
shall denote this (uniquely defined everywhere on $V$) sublinear
 function by the same symbol $\hat f$.

 Recall that simple functions $f$ and $g$ are {\it in general position} if 
$\hat f(x) \neq \widehat g(x)$ for all $x$ belonging to an open dense subset of $V$. 
In particular, generalized monomials are in  
general position whenever they are linearly independent.

Denote by $\mathit{Sim}(\cset^n)$ the set of all simple functions defined on $V$ and 
denote by $\mathit{Sim}_+(\cset^n)$ the set $\mathit{Sim}(\cset^n) \cap \maD_+(\cset^n)$. By 
$\mathit{Sbl}(V)$ denote the
 set of all (continuous) sublinear functions defined on $V = \rset^n$ and by
 $\mathit{Sbl}_+(V)$ denote the image $\widehat{\mathit{Sim}_+}(\cset^n)$ of $\mathit{Sim}_+(\cset^n)$ under the
 dequantization transform.

The following statements can be easily deduced from Theorem 1 and definitions.

\begin{cor}
The set $\mathit{Sim}_+(\cset^n)$ is a subsemiring of $\maD_+(\cset^n)$ and $\mathit{Sbl}_+(V)$
is an idempotent subsemiring of $\widehat{\maD_+}(V)$. The
dequantization transform generates an epimorphism of
$\mathit{Sim}_+(\cset^n)$ onto $\mathit{Sbl}_+(V)$. The set $\mathit{Sbl}(V)$ is an idempotent
semiring with respect to the operations
$(f\oplus g)(x) = \max \{ f(x), g(x)\}$, 
$(f\odot g)(x) = f(x) + g(x)$.
\end{cor}

\begin{cor}
Polynomials and generalized polynomials are simple functions.
\end{cor}

We shall say that functions $f, g\in\maD(V)$ are {\it asymptotically equivalent} 
whenever $\hat f = \widehat g$; any simple function $f$ is an {\it asymptotic 
monomial} whenever 
$\hat f$ is a linear function. A simple function $f$ will be called an {\it 
asymptotic polynomial} whenever $\hat f$ is a sum of a finite collection of 
nonequivalent asymptotic monomials.

\begin{cor}
Every asymptotic polynomial is a simple function.
\end{cor}

{\sc Example 1.} Generalized polynomials, logarithmic functions of 
(generalized) polynomials, and products of 
polynomials and logarithmic functions are asymptotic polynomials. This follows 
from our definitions and formula~(2).

\section{Subdifferentials of sublinear functions}

We shall use some elementary results from convex analysis. These results can be 
found, e.g., in \cite{MaTi2003}, ch. 1, \S 1.

For any function $p\in \mathit{Sbl}(V)$ we set
\begin{equation}
\partial p = \{\, v\in V\mid (v, x) \le p(x)\ \forall x\in V\,\}.
\end{equation}

It is well known from convex analysis that for any sublinear function $p$ the 
set $\partial p$ is exactly the {\it subdifferential} of $p$ at the origin. 
 The following propositions are also known in convex 
analysis.

\begin{prop}
Suppose $p_1,p_2\in \mathit{Sbl}(V)$, then
\begin{enumerate}
\item[1)] $\partial (p_1+p_2) = \partial p_1\odot\partial p_2 = \{\, v\in V\mid
v = v_1+v_2, \text{ where $v_1\in \partial p_1, v_2\in \partial p_2$}\,\}$;
\item[2)] $\partial (\max\{p_1(x), p_2(x)\}) = \partial p_1\oplus\partial p_2$.
\end{enumerate}
\end{prop}

Recall that $\partial p_1\oplus \partial p_2$ is a convex hull of the set 
$\partial p_1\cup \partial p_2$.

\begin{prop}
Suppose $p\in \mathit{Sbl}(V)$.  Then $\partial p$ is a nonempty convex compact 
subset of $V$.
\end{prop}

\begin{cor}
The map $p\mapsto \partial p$ is a homomorphism of the idempotent semiring
 $\mathit{Sbl}(V)$ (see Corollary 3) to the idempotent semiring $\mathcal{S}$ of all convex
 compact subsets of $V$ (see Section 1 above).
\end{cor}

\section{Newton sets for simple functions}

For any simple function $f\in \mathit{Sim}(\cset^n)$ let us denote by $N(f)$ the set
 $\partial(\hat f)$. We shall call $N(f)$ the {\it Newton set} of the
 function $f$.

\begin{prop}
For any simple function $f$, its Newton set $N(f)$ is a nonempty convex
 compact subset of $V$.
\end{prop}

This proposition follows from Proposition 3 and definitions.

\begin{thm}
Suppose that $f$ and $g$ are simple functions. Then
\begin{enumerate}
\item[1)] $N(fg) = N(f)\odot N(g) = \{\, v\in V\mid v = v_1 +v_2
 \text{ with
 $v_1 \in N(f), v_2 \in N(g)$}\,\}$;
\item[2)] $N(f+g) = N(f)\oplus N(g)$, if $f_1$ and $f_2$ are in general
 position or $f_1, f_2 \in \mathit{Sim}_+(\cset^n)$ {\rm (}recall that $N(f)\oplus N(g)$ 
is the convex hull of $N(f)\cup N(g)${\rm )}.
\end{enumerate}
\end{thm}

This theorem follows from Theorem~1, Proposition~2 and definitions.

\begin{cor}
The map $f\mapsto N(f)$ generates a homomorphism from $\mathit{Sim}_+(\cset^n)$ to 
$\mathcal{S}$.
\end{cor}

This statement follows from Theorem~1, Corollary~1, Corollary~6, and Theorem~2.

\begin{prop}
Let $f = m_{a,d}(x) = a \prod^n_{i=1} x_i^{d_i}$ be a monomial; here
$d = (d_1, \dots, d_n) \in V= \rset^n$ and $a$ is a nonzero
complex number. Then $N(f) = \{ d\}$.
\end{prop}
 
This follows from Proposition 1, Corollary 2 and definitions.

\begin{cor}
Let $f = \sum_{d\in D} m_{a_d,d}$ be a polynomial. Then $N(f)$ is the polytope 
$\oplus_{d\in D}\{d\}$, i.e.\ the convex hull of the finite set $D$.
\end{cor}

This statement follows from Theorem 2 and Proposition 5. Thus in this case
 $N(f)$ is the well-known classical Newton polytope of the polynomial $f$.

Now the following corollary is obvious.

\begin{cor}
Let $f$ be a generalized or asymptotic polynomial. Then its Newton set
 $N(f)$ is a convex polytope.
\end{cor}

{\sc Example 2}. Consider the one dimensional case, i.e., $V = \rset$ and 
suppose
 $f_1 = a_nx^n + a_{n-1}x^{n-1} + \dots + a_0$ and $f_2 = b_mx^m + b_{m-1}
 x^{m-1} + \dots + b_0$, where $a_n\neq 0$, $b_m\neq 0$, $a_0 \neq 0$,
 $b_0 \neq 0$. Then $N(f_1)$ is the segment $[0, n]$ and $N(f_2)$ is the
 segment $[0, m]$. So the map $f\mapsto N(f)$ corresponds to the map
 $f\mapsto \deg (f)$, where $\deg(f)$ is a degree of the polynomial $f$. In
 this case Theorem 2 means that $\deg(fg) = \deg f + \deg g$ and
 $\deg (f+g) = \max \{\deg f, \deg g\} = \max \{n, m\}$ if $a_i\geq 0$,
 $b_i\geq 0$ or $f$ and $g$ are in general position.

 {\sc Remark}.  The above results can be extended to the
 infinite-dimensional case.  This generalization will be the subject
 of another paper.


\begin{thebibliography}{99}

\bibitem{Mas87} V.~P.~Maslov,
{\it On a new superposition principle for 
optimization problems}, Uspekhi Mat. Nauk, [{\it Russian Math. 
Surveys\/}] {\bf 42} (1987), no.~3, 39--48.

\bibitem{LiMa95} G.~L.~Litvinov and V.~P.~Maslov, {\it 
Correspondence principle for idempotent calculus and some computer
applications}, (IHES/M/95/33), Institut des Hautes Etudes Scientifiques,
Bures-sur-Yvette, 1995. Also: \cite{Gun98}, p. 420--443 and 
arXiv:math.GM/0101021.

\bibitem{Gun98} J.~Gunawardena, (Ed.), {\it Idempotency}, Publ. of the 
Newton Institute, Vol. {\bf 11}, Cambridge University Press, Cambridge, 1998.

\bibitem{LiMaSh2001} G.~L.~Litvinov, V.~P.~Maslov, G.~B.~Shpiz,
{\it Idempotent functional analysis: an algebraic approach}, 
Mathematical Notes {\bf 69} (2001), no. 5, 
696--729. Also arXiv:math.FA/0009128.

\bibitem{MaTi2003} G.~G.~Magaril-Il'yaev and V.~M.~Tikhomirov,
{\it Convex analysis: theory and applications}, Translations of Mathematical 
Monographs, vol. {\bf 222}, Amer. Math. Soc., Providence, RI, 2003.

\bibitem{Ro70} R.~T.~Rockafellar, {\it Convex analysis}, Princeton 
Univ. Press, Princeton, 1970.

\end{thebibliography}
\end{document}